\documentclass[12pt]{article}
\usepackage{amsmath}
\usepackage{graphicx}
\usepackage{natbib}
\usepackage{mathtools}
\usepackage{appendix}
\usepackage{url} % not crucial - just used below for the URL 
\usepackage{hyperref}
\usepackage{comment}
\newcommand{\crim}{{\tt crim} }
\newcommand{\lstat}{{\tt lstat} }
\hypersetup{
colorlinks=true,
linkcolor=blue,
filecolor=magenta,      
urlcolor=cyan,
pdftitle={Overleaf Example},
pdfpagemode=FullScreen,
}

\newcommand{\blind}{0}

\newtheorem{definition}{Definition}
\newtheorem{theorem}{Theorem}
\DeclareMathOperator*{\argmin}{arg\,min}

\DeclarePairedDelimiter\ceil{\lceil}{\rceil}
\DeclarePairedDelimiter\floor{\lfloor}{\rfloor}

\begin{document}

\def\spacingset#1{\renewcommand{\baselinestretch}%
{#1}\small\normalsize} \spacingset{1}

%%%%%%%%%%%%%%%%%%%%%%%%%%%%%%%%%%%%%%%%%%%%%%%%%%%%%%%%%%%%%%%%%%%%%%%%%%%%%%

\if0\blind
{
  \title{\bf Varying Coefficient Model via Adaptive Spline Fitting}
  \author{Xufei Wang \\
    Two Sigma Investments, LP \thanks{The views expressed herein are the authors alone and are not necessarily the views of Two Sigma Investments, LP, or any of its affiliates.} \\
    Bo Jiang \\
    Two Sigma Investments, LP and \\
    Jun S. Liu \\
    Department of Statistics, Harvard University}
  \maketitle
} \fi

\if1\blind
{
  \bigskip
  \bigskip
  \bigskip
  \begin{center}
    {\LARGE\bf Title}
\end{center}
  \medskip
} \fi

\bigskip
\begin{abstract}
The varying coefficient model has received broad attention from researchers as it is a powerful dimension reduction tool for non-parametric modeling. Most existing varying coefficient models fitted with polynomial spline assume equidistant knots and take the number of knots as the hyperparameter. However, imposing equidistant knots appears to be too rigid, and determining the optimal number of knots systematically is also a challenge. In this article, we deal with this challenge by utilizing polynomial splines with adaptively selected and predictor-specific knots to fit the coefficients in varying coefficient models. An efficient dynamic programming algorithm is proposed to find the optimal solution. Numerical results show that the new method can achieve significantly smaller mean squared errors for coefficients compared with the equidistant spline fitting method.
\end{abstract}

\noindent%
\vfill

\newpage
\spacingset{1.5} % DON'T change the spacing!
\section{Introduction}
\label{sec:intro}

The problem of accurately estimating the relationship between a response variable and multiple predictor variables is fundamental to statistics and machine learning, and many other scientific applications. Among parametric models, the linear regression model is a simple and powerful approach, yet it is somewhat limited by its linearity assumption, which is often violated in real applications. In contrast, non-parametric models do not assume any specific type of relationship between the response variable and the predictors, which offer some flexibility in modeling nonlinear relationships, and are powerful alternatives to linear models. However, non-parametric models often need to impose certain local smoothness assumptions which are mostly achieved by employing a certain class of kernels or spline basis functions, to overcome the over-fitting issue. This is known to be plagued by the curse of dimensionality in that these methods are both ineffective in capturing the true relationship and computationally very expensive for high dimensional data sets. 

Serving as a bridge between linear and non-parametric models, the varying coefficient model \citep{Hastie:1993} provides an attractive compromise between simplicity and flexibility. In this class of models, the regression coefficients are not set constants and are allowed to depend on other conditioners. As a result, the varying coefficient model is more flexible because of the infinite dimensionality of the corresponding parameter spaces. Compared to standard non-parametric approaches, this method rises as a powerful strategy to cope with the curse of dimensionality. This method also inherits all advantages from linear models of being simple and interpretable. A typical setup for the varying coefficient model is as follows. Given response $y$ and predictors $X=(x_1,\ldots,x_p)^T$, the model assumes that $$y = \sum_{j=1}^p \beta_j(u) x_j+\epsilon,$$
where $u$ is the conditional random variable, which is usually a scalar. The varying coefficient model has a broad range of applications, including longitudinal data \citep{Huang:2004b, Tang:2012}, functional data \citep{Zhang:2014, Hu:2019}, and spatial data \citep{Wang:2019, Finley:2020}. Moreover, varying coefficient models could naturally be extended to time series contexts \citep{Huang:2004, Lin:2019}. 

There are three major approaches to estimate the coefficients $\beta_j(u)\  (j=1,\ldots,p)$. One widely acknowledged approach is the smoothing spline method proposed by \citet{Hastie:1993}, with a recent follow-up work using P-spline \citep{Jullion:2009}. \citet{Fan:1999} and \citet{Fan:2000} proposed a predictor-specific kernel approach for estimating the coefficients. The first step is to use local cubic smoothing to model function $\beta_j(u)$, and the second step re-applies local cubic smoothing on the residuals. A recent follow-up of their work is an adaptive estimator by \citet{Chen:2015}. Approximation of the coefficient functions using a basis expansion, e.g. polynomial B-splines, is always popular as it leads to a simple solution to estimation and inference with good theoretical properties. Compared with smoothing spline and kernel methods, the polynomial spline method with a finite number of knots strikes a balance between model flexibility and interpretability of the estimated coefficients. \citet{Huang:2002a}, \citet{Huang:2004}, and \citet{Huang:2004b} utilized a set of polynomial estimators. The assumptions made in these works are that the knots are equidistant and the number of knots is chosen such that the bias terms become asymptotically negligible to guarantee the local asymptotic normality.

Most polynomial spline approaches run optimization based on a set of finite-dimensional classes of functions, such as a space of polynomial B-splines with $L$ equally spaced knots. If the real turning points of the coefficients are not equidistant yet we still use the method assuming equally spaced knots, then $L$ should be large enough to make sure the distances between knots in the model are small enough to capture the resolution for the coefficients. %In theory, it is not clear how to systematically determine $L$ yet. 
In practice, $L$ is always chosen by a parameter searching process accompanied with other parameters, as people need to compare a set of parameters before determining the optimal fixed number of knots. Too few knots might ignore the high-frequency information of $\beta_j(u)$, while too many knots might over-fit the area where the coefficients barely change. Variable selection for varying coefficient models, especially when the number of predictors is larger than the sample size, is also an interesting direction to study.

In this paper, we propose two adaptive algorithms that are enabled by first fitting piece-wise linear functions with automatically selected turning points for univariate conditioner variable $u$. Compared with traditional methods above, these algorithms can automatically determine optimal positions of knots modeled as the turning points of the true coefficients. We prove that, if the coefficients are piece-wise linear in $u$, the selected knots by our methods are almost surely the true change points. We further combine the knots selection algorithms with the adaptive group LASSO method for variable selection in high-dimensional problems, following a similar idea as \citet{Wei:2011} who applies the adaptive group LASSO to basis expansions of predictors. Our simulation studies demonstrate that the new adaptive method achieves smaller mean squared errors for estimating the coefficients compared to available methods, and improves the variable selection performance. We finally apply the method to study the association between environmental factors and COVID-19 infected cases and observe time-varying effects of the environmental factors, as well as the Boston Housing data \citep{Harrison:1978}.

\section{Adaptive spline fitting methods}
\label{sec:methodology}
\subsection{Knots selection for polynomial spline}
\label{sec:global_knots}
In varying coefficient models, each coefficient $\beta_j(u)$ is a function of the conditional variable $u$, which can be estimated by fitting a polynomial spline on $u$. In this paper we assume that $u$ is a univariate variable. Let $X_i=(x_{i,1},\ldots,x_{i,p})^T\in\mathbf{R}^p$, $u_i$, and $y_i$ denote the $i$th observations of the predictor vector, the conditional variable, and the response variable, respectively, for $i=1,\ldots,n$. Suppose the knots are common to all coefficients located at $d_1<\ldots<d_L$, and the corresponding B-splines of degree $D$ are $B_k(u)\ (k=1,\ldots,D+L+1)$. Each varying coefficient can be represented as $\beta_j(u) = \sum_{k=1}^{D+L+1} c_{j,k}B_k(u)$, where the coefficients $c_{j,k}$ are estimated by minimizing the following sum of squared errors:
\begin{equation}
\label{equ:polynonialspline}
\sum_{i=1}^n \left\{y_i -\sum_{j=1}^p x_{i, j} \sum_{k=1}^{D+L+1} c_{j,k}B_k(u_i) \right\}^2.
\end{equation}
In previous work, the knots for polynomial spline were always chosen as equidistant quantiles of $u$ and are the same for all predictors. The approach is computationally straightforward, but the knots chosen cannot properly reflect the varying smoothness between and within the coefficients. We propose an adaptive knot selection approach in which the knots can be interpreted as turning points of the coefficients. 

For knots $d_1<\ldots<d_L$, we define the segmentation scheme $S=\{s_1,\ldots,s_{L+1}\}$ for the observed samples ordered by $u$, where $s_k=\{i\mid  d_k<u_i\leq d_{k+1}\}$, with $d_0=-\infty$ and $d_{L+1}=\infty$. If the true coefficients $\beta (u)=(\beta_1(u),\ldots,\beta_p(u))^T\in\mathbf{R}^p$ is a linear function of $u$ within each segment, i.e., $\beta(u)=a_s + b_s u$ for $a_s, b_s\in\mathbf{R}^p$, then the observed response satisfies 
\begin{equation}
\label{equ:piecewiselinear}
y=a_s^TX + b_s^T(uX) + \epsilon, \epsilon\sim N(0,\sigma_s^2).
\end{equation}
Thus, the coefficients can be estimated by maximizing the log-likelihood function, which is equivalent to minimizing the loss function:
\begin{equation}
\label{equ:loglikelihood}
L(S) = \sum_{s\in S} |s| \log \hat{\sigma}^2_s,
\end{equation}
where $\hat{\sigma}^2_s$ is the residual variance by regressing $y_i$ over $(X_i, u_iX_i)$ for $i\in s$.

Because any almost-everywhere continuous function can be approximated by piece-wise linear functions, we can employ the estimating framework in (\ref{equ:piecewiselinear}) and derive (\ref{equ:loglikelihood}). To avoid over-fitting when maximizing the log-likelihood over the parameters and segmentation schemes, we constrain that $|s|$ is greater than a lower bound $m_s=n^{-\alpha}\ (\alpha>0)$, and penalize the loss function with the number of segments
\begin{equation}
\label{equ:loglikelihood_penalized}
L(S, \lambda_0) = \sum_{s\in S} |s| \log \hat{\sigma}^2_s + \lambda_0 |S| \log(n),
\end{equation}
where $\lambda_0>0$ is the penalty strength. The resulting knots correspond to the segmentation scheme $S$ that minimizes the penalized loss function (\ref{equ:loglikelihood_penalized}). When $\lambda_0$ is very large, this strategy tends to select no knots, whereas when $\lambda_0$ gets close to 0 it can select as many knots as $n^{1-\alpha}$. We find the optimal $\lambda_0$ by minimizing the Bayesian information criterion \citep{Schwarz:1978} of the fitted model. For a given $\lambda_0$, suppose $L(\lambda_0)$ knots are finally proposed, and the fitted model is $\hat{f}(X,u)$. Then, we have
\begin{equation}
\label{equ:polynomialspline_BIC}
\textsc{bic}(\lambda_0) = n\log \left[\frac{1}{n} \sum_{i=1}^n \{y_i - f(X_i, u_i)\}^2 \right] + p\{L(\lambda_0)+D+1\} \log(n).
\end{equation}
The optimal $\lambda_0$ is selected by searching over a grid to minimize \textsc{bic}$(\lambda_0)$. We call this procedure the global adaptive knots selection strategy as it assumes that all the predictors have the same set of knots. We will discuss in Section~\ref{sec:individual_knots} how to allow each predictor to have its own set of knots.

Here we only use the piece-wise linear model \eqref{equ:piecewiselinear} and loss function \eqref{equ:loglikelihood_penalized} for knots selection, but will fit the varying coefficients with B-splines derived from the resulting knots via minimizing (\ref{equ:polynonialspline}). In this way, the fitted varying coefficients are smooth functions and the smoothness is determined by the degree of the splines. This method is referred to as the global adaptive spline fitting throughout the paper.

\subsection{Theoretical property of the selected knots}
The proposed method is invariant to marginal distribution of $u$. Without loss of generality, we assume that $u$ follows the uniform distribution in $[0,1]$. According to Definition~\ref{turningpoint} below, a turning point of $u$, denoted as $c_k$, is defined as a local maximum or minimum of any coefficient $\beta_i(u)\ (j=1,\ldots,p)$. In Theorem~\ref{theorem1}, we show that the adaptive knots selection approach can almost surely detect all the turning points of $\beta(u)$. 
\begin{definition}\label{turningpoint}
We call  $0<c_1<\ldots <c_K <1, K<\infty$ the turning points of $\beta(u)$ if for any $c_{k-1}<u_1<u_2<c_k<u_3<u_4<c_{k+1}\  (k=1,\ldots, K)$, 
\[\{\beta_j(u_1)-\beta_j(u_2)\}\{\beta_j(u_3)-\beta_j(u_4)\} < 0,\] 
for some index $j$ with $u_0=0,u_{K+1}=1$.
\end{definition}
\begin{theorem} \label{theorem1}
Suppose $y=\beta(u)^TX+\epsilon, u\sim \mathrm{Unif}(0,1), \epsilon \sim N(0,\sigma^2)$ where $X$ is bounded and $\beta(u)$ is a bounded continuous function. Moreover, assume $X,\ u,\ \epsilon$ are independent. Let $0<c_1<\ldots <c_K <1$ be the turning points in Definition~\ref{turningpoint} and $d_1<\ldots <d_L$ be the selected knots with $m_s=n^{-\alpha}\  (\alpha > 0)$, then for $0<\gamma<1-\alpha$ and $\lambda_0$ large enough,
\begin{eqnarray*}
\mathrm{pr}\left(L\geq K, \max_{k=1}^K \min_{l=1}^L |d_l-c_k| < n^{-\gamma}\right) \to 1,\quad n\to\infty.
\end{eqnarray*}
\end{theorem}
Although the adaptive spline fitting method is motivated by a piece-wise linear model, Theorem 1 shows that with probability approaching 1, we can detect all the turning points accurately under general varying coefficient functions. The selected knots could be a superset of the real turning points especially when $\lambda_0$ is small, so we tune $\lambda_0$ with \textsc{bic} (\ref{equ:polynomialspline_BIC}) which penalizes the number of selected knots, to find the optimal set.

When the varying coefficient functions are piece-wise linear, each coefficient $\beta_j(u)\ (j=1,\ldots,p)$ can be almost surely defined as 
\begin{eqnarray*}
\beta_j(u) = a_{j,k} + b_{j,k} u,\quad c_{j,k-1} < u \leq c_{j, k},\quad k=1,\ldots, K_j+1,
\end{eqnarray*}
where $c_{j,0}=0$, $c_{j,K_j+1}=1$ and $(a_{j,k}-a_{j,k+1})^2 + (b_{j,k}-b_{j,k+1})^2 > 0\ (k=1,\ldots,K_j)$ if $K_j \geq 1$. When $K_j=0$ the varying coefficient $\beta_j(u)$ is a simple linear function of $u$, otherwise we call $c_{j,k}\ (k=1,\ldots,K_j)$ the change points for coefficients $\beta_j(u)$, because the linear relationship varies before and after these points. Then $\bigcup_{K_j \geq 1} \{c_{j,1},\ldots,c_{j,K_j}\}$ are all the change points of the entire varying coefficient function. In Theorem~\ref{theorem2}, we show that the adaptive knots selection method can almost surely discover the change points of $\beta(u)$ without false positive selection.
 
\begin{theorem} \label{theorem2}
Suppose $(X,y,u)$ follow the same assumption as in Theorem~\ref{theorem1}, and $\beta(u)$ is a piece-wise linear function of $u$. Let $0<c_1<\ldots <c_K <1$ be the change points defined as above and $d_1<\ldots <d_L$ be the selected knots with $m_s=n^{-\alpha}\ (\alpha > 0)$, then for $0<\gamma<1-\alpha$ and $\lambda_0$ large enough, 
\begin{eqnarray*}
\mathrm{pr}\left(L= K, \max_{k=1}^K |d_k-c_k| < n^{-\gamma} \right) \to 1,\quad n\to\infty.
\end{eqnarray*}
\end{theorem}
The theorem shows that if the varying coefficient function is piece-wise linear, the method can discover all the change points with almost 100\% accuracy.

\subsection{Dynamic programming algorithm for adaptive knots selection}
The brute force algorithm to compute the optimal knots has a computation complexity of $O(2^n)$ and is impractical. As summarized by the following algorithm, we propose a dynamic programming approach whose computation complexity is of order $O(n^2)$. If we further assume that the knots can only be chosen from a predetermined set $\mathcal{M}$, e.g. $\frac{m}{\sqrt{n}}\ (m=1,\ldots,\floor{\sqrt{n}}-1)$ quantiles of $u_i$'s, the computation complexity can be further reduced to $O(|\mathcal{M}|^2)$. Note that the algorithm in Section~2.4 of \cite{Wang:2017} is a special case with a constant $x$. The algorithm is summarized in the following three steps:

\begin{enumerate}
\item Data preparation: Arrange the data $(X_i,u_i,y_i)\ (i=1,\ldots,n)$ according to order of $u_i$ from small to large. Without lose of generality, we assume that $u_1<\cdots<u_n$.
\item \label{step:knotselection}Obtain the minimum loss by forward recursion: Define $m_s=\ceil{n^{-\alpha}}\ (\alpha >0)$ as the smallest segment size, and set $\lambda=\lambda_0\log(n)$. Initialize two sequence: $(\mathrm{Loss}_i,\mathrm{Prev}_i)\  (i=1,\ldots n)$ with $\mathrm{Loss}_0=0$. For $i=m_s,\ldots,n$, recursively fill in entries of the tables with
\begin{equation*}
\mathrm{Loss}_i = \min_{i'\in I_i} (\mathrm{Loss}_{i'-1} + l_{i':i} + \lambda),\quad \mathrm{Prev}_i = \argmin_{i'\in I_i} (\mathrm{Loss}_{i'-1} + l_{i':i} + \lambda).
\end{equation*}
Here $I_i=\{1\}\cup\{m_s+1,\ldots,i-m_s+1\}$ and $l_{i':i}=(i-i'+1)\log\hat{\sigma}^2_{i':i}$ where $\hat{\sigma}^2_{i':i}$ is the residual variance of regressing $y_k$ on $(X_k, u_kX_k)\ (k=i',\ldots,i)$.
\item \label{step:final} Determine knots by backward tracing: Let $p=\mathrm{Prev}_n$. If $p=1$ no knot is  needed; otherwise, we recursively add $0.5(u_{p-1}+u_p)$ as a new knot with $p=\mathrm{Prev}_p$ until $p=1$.
\end{enumerate}

When the algorithm is run with a grid of $\lambda_0$, we repeat Step~\ref{step:knotselection} and~\ref{step:final} for all the $\lambda_0$'s, and return the final model with the minimum \textsc{bic}.

\subsection{Predictor specific adaptive knots selection}
\label{sec:individual_knots}
The global adaptive knots selection method introduced in Section~\ref{sec:global_knots} assumes that the set of knot locations are the same for all predictors, similar to most of the literature for polynomial spline fitting. However, different coefficients may have a different resolution of smoothness relative to $u$, and a different set of knots for each predictor is preferable. Here we propose a predictor-specific adaptive spline fitting algorithm on top of the global knot selection. Suppose the fitted model for the global adaptive spline fitting is $\hat{f}(X,u)=\sum_{j=1}^p \hat{\beta}_j(u)x_j$. For predictor $j$, the knots can be updated by performing the same knots selection procedure between it and the residual without it, that is 
\begin{equation}
\label{equ:residual}
r_i = y_i - \sum_{j'\neq j} \hat{\beta}_{j'}(u_i) x_{i,j'},\quad i=1,\ldots,n.
\end{equation}
If \textsc{bic} in~(\ref{equ:polynomialspline_BIC}) is smaller for the corresponding new polynomial spline model with the new knots, the knots and fitted model are update. The steps are repeated until \textsc{bic} cannot be further minimized. The following three steps summarizes the algorithm:
\begin{enumerate}
\item \label{step:initial} global adaptive spline: Run the global adaptive spline fitting algorithm for $(X_i,u_i,y_i)\ (i=1,\ldots,n)$. Denote $\hat{f}(X,u)$ as the fitted model and compute model \textsc{bic} with (\ref{equ:polynomialspline_BIC}).
\item Update knots with \textsc{bic}: For $j=1,\dots,p$, compute residual variable $r_i$ by (\ref{equ:residual}). Run the global adaptive spline fitting algorithm for the residual and predictor $j$, that is $(x_{i,j},u_i,r_i)\ (i=1,\ldots,n)$. Denote the new fitted model as $\hat{f}^j(X,u)$ and corresponding \textsc{bic} as $\textsc{bic}_j$. Note that when we assume different number of knots for each predictor, the term $p\{L(\lambda_0)+D+1\} \log(n)$  in \eqref{equ:polynomialspline_BIC} 
should be replaced by $\left\{\sum_{j=1}^p L_j(\lambda_0) + p(D+1)\right\} \log(n)$, where $L_j(\lambda_0)$ is the number of knots for predictor $j$.
\item \label{step:update} Recursively update model: If $\min \textsc{bic}_j<\textsc{bic}$ and $j^*=\argmin \textsc{bic}_j$, update the current model by $\hat{f}(X,u)=\hat{f}^{j^*}(X,u)$, and repeat Step~\ref{step:update}. Otherwise return the fitted model $\hat{f}(X,u)$.
\end{enumerate}
An alternative approach to model the heterogeneous relationships between coefficients is to replace the initial model in Step~\ref{step:initial} with $\hat{f}(X,u)=0$, and repeat the following two steps. However, starting from the global model is preferred because fitting to residual instead of the original response minimizes the mean squared error more efficiently. Simulation studies in Section~\ref{sec:simulation1} show that the predictor-specific knots can further reduce mean squared error for the fitted coefficient, compared with the global knot selection approach. 

\subsection{Knots selection in sparse high-dimensional problems}
When the number of predictors is large and the number of predictors with non-zero varying coefficients is small, we perform a variable selection for all the predictors. For predictor $j$, we run the knots selection procedure between it and the response, and construct B-spline functions as $\{B_{j,k}(u)\}\ (L=1,\ldots,L_j+D+1)$ where $L_j$ is the number of knots and $D$ is the degree of the B-splines. Then, we perform the variable selection method for varying coefficient model proposed by \citet{Wei:2011}, which is a generalization of group LASSO \citep{Yuan:2006} and adaptive LASSO \citep{Zou:2006}. \citet{Wei:2011} shows that group LASSO tends to over select variables and suggests adaptive group LASSO for variable selection. In their original algorithm, the knots for each predictor are chosen as equidistant quantiles and not predictor-specific. The following three steps summarize our proposed algorithm:
\begin{enumerate}
\item \label{step:marginal} Select knots for each predictor: Run the global adaptive spline fitting algorithm between each predictor $x_{i,j}$ and response $y_i$. Denote the corresponding B-splines as $B_{j,k}(u)\ (k=1,\ldots,L_j+D+1)$ where $L_j$ is the number of knots for predictor $j$ and $D$ is degree of splines.
\item \label{step:groupLASSO} Group LASSO for first step variable selection: Run group LASSO algorithm for the following loss function
\begin{eqnarray*}
\frac{1}{n} \sum_{i=1}^n \left\{y_i - \sum_{j=1}^p x_{i,j} \sum_{k=1}^{L_j+D+1} c_{j,k} B_{j,k}(u_i)\right\} + \lambda_1 \sum_{j=1}^p (c_j'R_jc_j)^{1/2},\quad  \lambda_1>0
\end{eqnarray*}
where $c_j=(c_{j,1},\ldots,c_{j,L_j+D+1})$ and $R_j$ is the kernel matrix whose $(k_1,k_2)$ element is $E\{B_{j,k_1}(u) *  B_{j,k_2}(u)\}$. Denote the fitted coefficients as $\tilde{c}_{j,k}$.
\item \label{step:adaptivegroupLASSO} Adaptive group LASSO for second step variable selection: Run adaptive group LASSO algorithm for the updated loss function 
\begin{eqnarray*}
\frac{1}{n} \sum_{i=1}^n \left\{y_i - \sum_{j=1}^p x_{i,j} \sum_{k=1}^{L_j+D+1} c_{j,k} B_{j,k}(u_i)\right\} + \lambda_2 \sum_{j=1}^p w_j (c_j'R_jc_j)^{1/2},\quad \lambda_2 > 0
\end{eqnarray*}
with weight as $w_k=\infty$ if $\tilde{c}_j'R_j\tilde{c}_j=0$, otherwise $w_k=(\tilde{c}_j'R_j\tilde{c}_j)^{-1/2}$. Denote the fitted coefficients as $\hat{c}_{j,k}$ and the selected variables are those with $\hat{c}_j'R_j\hat{c}_j \neq 0$.
\end{enumerate}
For Steps~\ref{step:groupLASSO} and~\ref{step:adaptivegroupLASSO}, parameter $\lambda_1$ and $\lambda_2$ are chosen by minimizing \textsc{bic} (\ref{equ:polynomialspline_BIC}) for the fitted model, where the degree of freedom is computed with only the selected predictors. Simulation studies in Section~\ref{sec:simulation2} show that, with the knots selection Step~\ref{step:marginal}, the algorithm can always choose the correct predictors with a reasonable number of samples.

\section{Empirical studies}
\label{sec:simulation1}
\subsection{Simulation study for adaptive spline fitting}
\begin{figure}[t]
\centering
\includegraphics[width=0.9\textwidth]{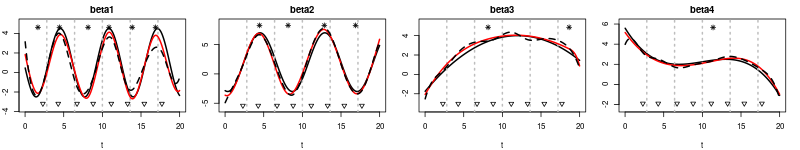}
\caption{The true coefficients and fitted coefficients with equidistant and predictor-specific knots. Black line: true coefficients; triangle: equidistant knots; dashed lines: fitted coefficients with equidistant knots; stars: predictor-specific knots; red: fitted coefficients with predictor-specific knots.} 
\label{fig:simulation}
\end{figure}

In this section, we compare the global and predictor-specific adaptive spline fitting approaches we introduced with the equidistant spline fitting approach as well as a commonly used kernel method package \texttt{tvReg} by \citet{Casas:2019}, using the simulation examples from \citet{Tang:2012}. The model simulates a longitudinal data, which are frequently encountered in biomedical applications. In this example, the conditional random variable represents time and for convenience, we use notation $t$ instead of $u$. We simulated $n$ individuals, each having a scheduled time set $\{0,1,\ldots,19\}$ to generate observations. A scheduled time can be skipped with probability 0.6, so no observation will be generated at the skipped time. For each non-skipped scheduled time, the real observed time is the scheduled time plus a random disturbance from $\mathrm{Unif}(0,1)$. Consider the time-dependent predictors $X(t)=(x_1(t),x_2(t),x_3(t),x_4(t))^T$ with
\begin{eqnarray*}
x_1(t) =1, && x_2(t) \sim \mathrm{Bern}(0.6), \\
x_3(t) \sim \mathrm{Unif}(0.1t, 2+0.1t), && x_4(t) ~\mid ~x_3(t) \sim N\left(0, \frac{1+x_3(t)}{2+x_3(t)}\right).
\end{eqnarray*}
The response for individual $i$ at observed time $t_{i,k}$ is $y_i(t_{i,k}) = \sum_{j=1}^4 \beta_j(t_{i,k}) x_{i,j}(t_{i,k}) + \epsilon_i(t_{i,k})$, with 
\begin{eqnarray*}
\beta_1(t) = 1 + 3.5 \sin(t-3), &&\beta_2(t) = 2 - 5 \cos (0.75t-0.25), \\
\beta_3(t) = 4 - 0.04(t-12)^2, && \beta_4(t) = 1 + 0.125t + 4.6\left(1 - 0.1t\right)^3.
\end{eqnarray*}
The random error $\epsilon_i(t_{i,k})$ are independent from the predictors, and generated with two components $$\epsilon_i(t_{i,k})=v_i(t_{i,k})+e_i(t_{i,k}), $$
where $e_i(t_{i,k})\stackrel{i.i.d.}{\sim} N(0,4)$, and $v_i(t_{i,k}) \sim N(0, 4)$ with correlation structure $$\mathrm{cor}\left\{v_{i_1}(t_{i_1,k_1}), v_{i_2}(t_{i_2,k_2})\right\} = I(i_1=i_2) \exp(-|t_{i_1,k_1}-t_{i_2,k_2}|).$$
Random errors are positively correlated within the same individual and are independent between different individuals. Figure~\ref{fig:simulation} shows the true coefficients and the fitted coefficients by the equidistant and predictor-specific spline fitting methods for an example with $n=200$, as well as the selected knots. Note that for the equidistant fitting approach, the number of knots is also determined by minimizing the model \textsc{bic} (\ref{equ:polynomialspline_BIC}). We observe that the fitted coefficients by the predictor-specific method are smoother than those by the equidistant method, especially for the less volatile coefficients $\beta_3(t)$ and $\beta_4(t)$. This is because the predictor-specific method uses only 1 knot for these two coefficient which better reflects the resolution, while the equidistant method uses 7 knots.

We compare the three methods by  the mean square errors of their estimated coefficients, i.e., 
\begin{equation}
\label{equ:msebeta}
\textsc{mse}_j =  \frac{1}{N} \sum_{i=1}^n \sum_{k=1}^{n_i} \frac{\left\{\hat{\beta}_j(t_{i,k}) - \beta_j(t_{i,k}) \right\}^2}{\mathrm{range}(\beta_j)^2},
\end{equation}
where ${\beta}_j(t)$ and $\hat{\beta}_j(t)$ are the true and estimated coefficients, respectively, $n_i$ is the total number of observations for individual $i$ and $N=\sum_{i=1}^n n_i$. We run the simulation with $n=200$ for 1000 repetitions. For adaptive spline methods we only consider knots from $\frac{m}{\sqrt{N}}\  (m=1,\ldots,\floor{\sqrt{N}})$ quantiles of $t_{i,k}$.

\begin{table}[t]
\centering
\begin{tabular}{c| c c c c}
& $\textsc{mse}_1 * 1e2$ & $\textsc{mse}_2* 1e2$ & $\textsc{mse}_3* 1e2$ & $\textsc{mse}_4* 1e2$\\
\hline
equidistant & 2.53 (1.14)& 0.26 (0.10) & 0.64 (0.29) & 0.09 (0.04) \\
global & 2.24 (1.12) & 0.32 (0.12) & 0.55 (0.26) & 0.08 (0.04)\\
predictor-specific & \textbf{1.30} (0.83) & \textbf{0.23} (0.10) & \textbf{0.28} (0.21) & \textbf{0.04} (0.03) \\
kernel & 2.80 (0.92) & 0.38 (0.11) & 0.81 (0.26) &  0.14 (0.04) \\
\end{tabular}
\caption{$\textsc{mse}_j$ for the global and predictor-specific adaptive spline fitting methods, compared with kernel method.}
\label{tab:MSE}
\end{table}
Table~\ref{tab:MSE} shows the average $\textsc{mse}_j$ for the proposed global and predictor-specific methods, in comparison with the equidistant spline fitting and the kernel method. We find that the predictor-specific method has the smallest $\textsc{mse}_j$ for all four coefficients, significantly outperforming the other two methods. Interestingly, the equidistant method selects on average 6.9 knots, the global adaptive method selects on average 6.1 global knots for all predictors, whereas the predictor-specific method selects fewer on average: 5.8 knots for $x_1(t)$, 4.2 for $x_2(t)$, 1.5 for $x_3(t)$ and 0.8 for $x_4(t)$. The result is expected since $\beta_3(t)$ and $\beta_4(t)$ are less volatile than $\beta_1(t)$ and $\beta_2(t)$. It appears that the predictor-specific procedure can use a smaller number of knots to achieve a better fitting than the other two methods.

\subsection{Simulation study for variable selection}
\label{sec:simulation2}
We use the simulation example from \citet{Wei:2011} to compare the performance of our method with one using adaptive group LASSO and equidistant knots. Similar to the previous subsection, there are $n$ individuals, each having a scheduled time set $\{0,1,\ldots,29\}$ to generate observations and a skipping probability of 0.6. For each non-skipped scheduled time, the real observed time is the scheduled time adding a random disturbance generated from $\mathrm{Unif}(0,1)$. We construct $p=500$ time-dependent predictors as follows:
\begin{eqnarray*}
x_1(t) \sim \mathrm{Unif}\left(0.05+0.1t,2.05+0.1t\right),&\ &x_j(t) ~\mid~ x_1(t) \sim N\left(0,\frac{1+x_1(t)}{2+x_1(t)}\right),\quad j=2,\ldots, 5\\
x_6(t) \sim N\left(3\exp\{(t+0.5)/30\},1\right) &\ & x_j(t) \stackrel{i.i.d.} \sim N(0, 4),\quad j=7,\ldots,500.
\end{eqnarray*}
The same individual's predictors $x_j(t)\ (j=7,\ldots,500)$ are correlated with  $\mathrm{cor}\{x_j(t), x_j(s)\}=\exp(-|t-s|)$. The response for individual $i$ at observed time $t_{i,k}$ is $y_i(t_{i,k})=\sum_{j=1}^6 \beta_j(t_{i,k}) x_{i,j}(t_{i,k}) + \epsilon_i(t_{i,k})$. The time-varying coefficients $\beta_j(t)\ (j=1,\ldots,6)$ are
\begin{eqnarray*}
\beta_1(t) = 15 + 20\sin\{\pi (t+0.5)/15\},&\ & \beta_2(t) = 15 + 20 \cos \{\pi (t+0.5)/15\}, \\
\beta_3(t) = 2 - 3\sin\{\pi(t-24.5)/15\}, &\ &\beta_4(t) = 2 - 3\cos\{\pi(t-24.5)/15\},\\
\beta_5(t) = 6-0.2(t+0.5)^2, &\ & \beta_6(t) = -4 + 5*10^{-4}(19.5-t)^3.
\end{eqnarray*}
The random error $\epsilon_i(t_{i,k})$ is independent from the predictors and follows the same distribution as that in Section~\ref{sec:simulation1}. We simulate cases with $n=50,100,200$ and replicate each set for 200 times. Three metrics are considered: average number of selected variables, percentage of cases when there is no false negative, and percentage of cases when there is no false positive or negative. A comparison of our method with the variable selection method using equidistant knots \citep{Wei:2011} is summarized in Table~\ref{tab:metric}. We find that our method clearly outperforms the method with no predictor-specific knots selection, and can always select the correct predictors when $n$ is larger than 100.

\begin{table}[t]
\begin{tabular}{c | c c c }
\multicolumn{4}{c}{equidistant knots}  \\
& \# selected variables & \% no false negative  & \% no false positive or negative \\
\hline 
$n=50$ & 7.04 & 72 & 68   \\
$n=100$& 6.21  & 87  & 84  \\
$n=200$& 6.13  & 99  & 93  \\
\multicolumn{4}{c}{adaptive selected knots}  \\
& \# selected variables & \% no false negative  & \% no false positive or negative \\
\hline 
$n=50$ & 5.96  & 96.50 &96.50 \\
$n=100$&  6.00 & 100 & 100 \\
$n=200$& 6.00 & 100 & 100 \\
\end{tabular}
\caption{Variable selection performance for adaptive group LASSO with and without predictor-dependent knots selection.}
\label{tab:metric}
\end{table}

\section{Applications}

\subsection{ Environmental factors and COVID-19}
\begin{figure}[t]
\centering
\includegraphics[width=0.9\textwidth]{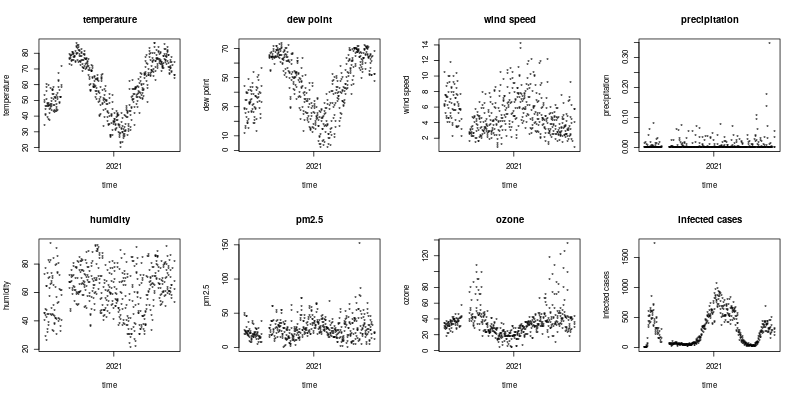}
\caption{Environmental measurements and COVID-19 infected cases in New York County, NY.}
\label{fig:environment}
\end{figure}

The dataset we investigate contains daily measurements of meteorological data and air quality data in 7 counties of the state of New York between March 1, 2020, and September 30, 2021. The meteorological data were obtained from the National Oceanic and Atmospheric Administration Regional Climate Centers, Northeast Regional Climate Center at Cornell University: \url{http://www.nrcc.cornell.edu}. The daily data are based on the average of the hourly measurements of several stations in each county and composed of records of five meteorological components: temperature in Fahrenheit, dew point in Fahrenheit, wind speed in miles per hour, precipitation in inches, and humidity in percentage. The air quality data were from the Environmental Protection Agency: \url{https://www.epa.gov}. The data contain daily records of two major air quality components, the fine particles with an aerodynamic diameter of 2.5$\rm{\mu m}$ or less, i.e., $\rm{PM_{2.5}}$, in $\rm{\mu g/m^3}$ and ozone in $\rm{\mu g/m^3}$. The objective of the study is to understand the association between the meteorological measurements, in conjunction with pollutant levels, and the number of infected cases for COVID-19, a contagious disease caused by severe acute respiratory syndrome coronavirus 2, and examine whether the association varies over time. The daily infected records were retrieved from the official website of the Department of Health, New York state: \url{https://data.ny.gov}. Figure~\ref{fig:environment} shows scatter plots of daily infected cases in New York County and the 7 environmental components over time. In order to remove the variation of recorded cases between weekdays and weekends, in the following study, we consider the weekly averaged infected cases which are the average between each day and the following 6 days. We also find that the temperature factor and dew point factor are highly correlated, and we remove the dew point factor when fitting the model.

We first take the logarithmic transformation of the weekly averaged infected cases, denoted as $y$, and then fit a varying coefficient model with the following predictors: $x_1=1$ as intercept, $x_2$ as temperature, $x_3$ as wind speed, $x_4$ as precipitation, $x_5$ as humidity, $x_6$ as $\rm{PM_{2.5}}$ and $x_7$ as ozone. Time $t$ is the conditioner for our model. All predictors except the constant are normalized to make varying coefficients $\beta_j(t)$ comparable. The varying coefficient model has the form:
\begin{equation}
\label{equ:environment}
y_i(t_{i,k}) = \sum_{j=1}^7 \beta_j(t_{i,k})x_{i,j}(t_{i,k}) + \epsilon(t_{i,k}),
\end{equation}
where $t_{i,k}$ is the $k$th record time for the $i$th county, with $y_i(t_{i,k})$ and $x_{i,j}(t_{i,k})$ being the corresponding records for county $i$ at time $t_{i,k}$, and $\epsilon(t_{i,k})\stackrel{iid}{\sim} N(0,\sigma^2)$ denotes the error term.

\begin{figure}[t]
\centering
\includegraphics[width=0.9\textwidth]{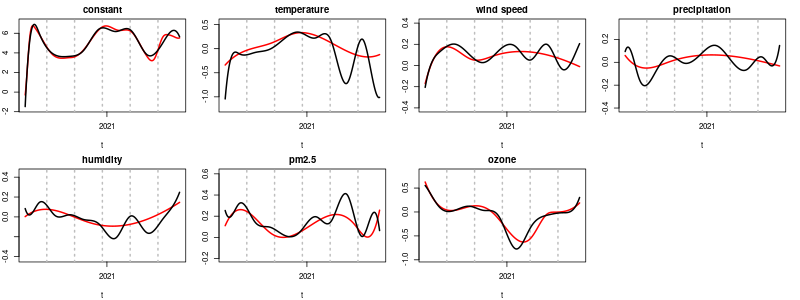}
\caption{Black lines: fitted coefficients with predictor-specific knots; red lines: fitted coefficients with equidistant knots.}
\label{fig:coefficients}
\end{figure}

We apply the equidistant and the proposed predictor-specific adaptive spline fitting method to fit the data. Figure~\ref{fig:coefficients} shows the fitted $\beta_j(t)$ for each predictor by the two methods. Figures show that there is a strong time effect on each coefficient function. For example, there are several peaks for the intercept, which correspond to the initial outbreak and delta variant outbreak. Moreover, there are rapid changes of coefficients around March 2020. This could be explained by the fact that at the beginning of the outbreak, the test cases were fewer and the number of infected cases was underestimated. Moreover, the coefficient curves show that the most important predictor is temperature. For most of the period, the coefficient is negative, indicating that high temperature is negatively associated with transmission of the virus, which is also observed in the study by \citet{Notari:2021} such that COVID-19 spread is slower at high temperatures. Besides the negatively correlated relationship, our analysis also demonstrates the time-varying property of the coefficient. Moreover, the fitted coefficients by the predictor-specific knots are less volatile than the equidistant knots, especially for temperature, wind speed, precipitation, humidity and $\rm{PM_{2.5}}$.

We evaluate the predictivity of proposed approach by a rolling window approach, with the training size of at least of 1 year, and rolling window of 1 week. That is to say for each date $t$ after March 1, 2021, we fit two models with equidistant and predictor-specific knots with $\left(x_{i,j}(t_{i,j}), y(t_{i,j})\right)\ (t_{i,j} < t)$, and predicts $y(t_{i,j})\ (t\leq t_{i,j} < t+7)$. The root mean squared error for equidistant knots fitting is 0.932 and for predictor-specific knots fitting is 0.901.

\begin{figure}[t]
\centering
\includegraphics[width=0.9\textwidth]{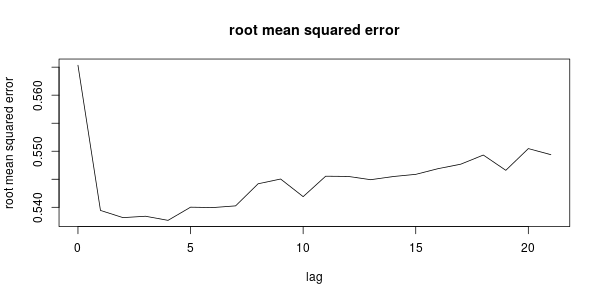}
\caption{Root mean squared error for logarithm of weekly averaged infected cases, with lag $\tau=0,\ldots,21$ days.}
\label{fig:lag}
\end{figure}

We note that environmental factors may not have an immediate effect on the number of recorded COVID-19 cases since there is 2-14 days of incubation period before the onset of symptoms and it takes another 2-7 days before a test result is available. To study whether there are lagging effects between the predictor and response variables, we fit a varying coefficient model with predictor-specific knots, for each time lag $\tau$, i.e., using data $y_i(t_{i,k} + \tau)$ and $x_{i,j}(t_{i,k})\ (j=1,\ldots,7)$, to fit model (\ref{equ:environment}) similarly as in \cite{Fan:1999}. Figure~\ref{fig:lag} shows the residual root mean squared error for each time lag. We find that the root mean squared error is the smallest with 4 days lag. Note that $y_i(t)$ represents the logarithm of the weekly average between day $t$ and $t+6$. This means the predictors at day $t$ are most predictive for infected numbers between day $t+4$ and $t+10$, which is consistent with the incubation period and test duration. %Figure~\ref{fig:coefficients} also show the lagged coefficients with $\tau=4$ in dashes, revealing nearly identical shapes with the non-lagged coefficients except the coefficient for $\rm{PM_{2.5}}$, which may indicate a large day-to-day variation of the $\rm{PM_{2.5}}$ measurements. 

\subsection{The Boston housing data}
We consider the Boston housing price data from \cite{Harrison:1978} with $n=506$ observations for the census districts of the Boston metropolitan area. The data is available in the R-package \texttt{lmbench}. Following \citep{Wang:2009, Hu:2012}, we use \texttt{medv} (mediam value of owner-occupied homes in 1,000 USD) as the response, and \lstat as the conditioner, which is defined as a linear combination of the proportion of adults with high school education or above and proportion of 
%and 1/2 proportion of 
male workers classified as laborers. 
The predictors are: {\tt int} (the intercept), \crim  (per capita crime rate by town), {\tt rm} (average number of rooms per dwelling), {\tt ptratio} (pupil-teacher ratio by town), \texttt{nox} (nitric oxides concentration parts per 10 million), {\tt tax} (full-value property-tax rate per 10,000 USD), and {\tt age} (proportion of owner-occupied units built prior to 1940). We transform the conditioner \lstat so that its marginal distribution is $\rm{U}(0,1)$, and make a logarithm transformation of the response \texttt{medv}. For the predictors (except for intercept), we first transform them to the standard normal. Since some of the predictors are highly correlated with the conditioner, we further
%so their marginal distributions are $\rm{N}(0,1)$, them we 
regress each predictor against the transformed \lstat, and use the normalized residuals in the follow-up analysis. 
%We perform the regression because we find the predictors and the conditioner are highly correlated. 

\begin{figure}[t]
\centering
\includegraphics[width=1.0\textwidth]{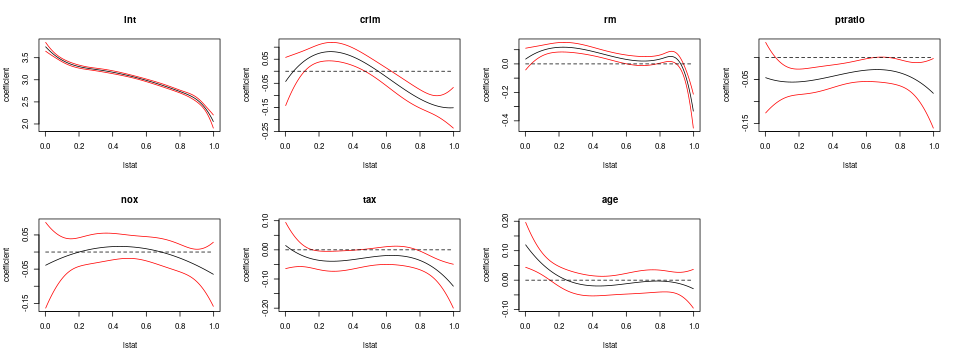}
\caption{Fitted coefficient for Boston Housing data, with conditioner as lstat.}
\label{fig:boston_housing}
\end{figure}

We fit a %apply the 
predictor-specific 
varying coefficient linear model to predict the response with the residualized predictors using \lstat as the conditioner. Figure~\ref{fig:boston_housing} shows the fitted coefficients as a function of the conditioner, with the red lines showing the 95\% confidence interval and the dashed lines being the x-axis. The confidence interval are computed conditioning on the selected knots for each predictor. The results show clear conditioner-varying effects of most predictors. 
%Given the range of the coefficients, we 
The intercept appears to vary most significantly along with {\tt lstat}, demonstrating that the
%and it is negatively correlated with lstat, so
the housing price is  negatively and nearly linearly impacted by {\tt lstat}. The coefficient for {\tt rm} is in general positive, but trends towards insignificant as \lstat increases. An interpretation is that houses with more rooms  are generally more expensive, but its impact becomes insignificant in areas where \lstat is large. The variable \crim is highly correlated with {\tt lstat}. After regressing out {\tt lstat}, the residualized \crim has an interesting fluctuation: varying from insignificant to positive and then to negative. We suspect that the positive impact of \crim for certain areas may be due to a confounding effect with other unused variables such as the location's convenience and attractiveness for tourists.
%Besides, the coefficient is close to zero when lstat is large, which shows that for area with big lstat, the effect of number of rooms for housing price is small. 
%Besides, the coefficient for age is always negative except when lstat is smaller than 0.225. This is expected because new houses tends to be more expensive.

We also compare the predictive power of the simple linear model and the varying coefficient model by 10-fold cross validation. For the simple linear model, we include all the predictors used in the varying coefficient model as well as the conditioner \lstat. The mean squared error (MSE) for the simple linear model is 23.52 %0.235 
%root mean squared error for simple linear model is 0.485, while the root mean squared error for varying coefficient model is 0.455
while the MSE for the varying coefficient model is 20.51. %0.207. 
When comparing the MSE, the observed and predicted responses are transformed back to the original scale. The results show that the varying coefficient model can better describe the relationship between the housing price and the predictors.

\section{Discussion}
\label{sec:conclusion}
In this paper, we introduce two algorithms for fitting varying coefficient models with adaptive polynomial splines. The first algorithm selects knots globally using a recursive method, assuming the same set of knots for all the predictors, whereas the second algorithm allows each predictor to have its own set of knots and also uses a recursive method for individualized knots selection. 

Coefficients modeled by polynomial splines with a finite number of non-regularly positioned knots are more flexible as well as more interpretable than the standard ones with equidistant knot placements. Simulation studies show that both global and predictor-specific algorithms outperform the commonly used kernel method, as well as equidistant spline fitting method, in terms of mean squared errors, while the predictor-specific algorithm achieves the best performance. A fast dynamic programming algorithm is introduced, with a computation complexity no more than $O(n^2)$ and can be of order $O(n)$ if the knots are chosen from $\frac{m}{\sqrt{n}}\ (m=1,\ldots,\floor{\sqrt{n}}-1)$ quantiles of the conditional variable $u$. 

Throughout the paper we assume that the conditioner variable $u$ is univariate. The proposed predictor-specific spline approach can be easily generalized to the case when each coefficient $\beta_j(u)$ has its own univariate conditioner variable $u$. However, it is still a challenging problem to generalize the proposed method to multi-dimensional conditioners and to model correlated errors.

\section*{Acknowledgement}
We are grateful to the National Oceanic and Atmospheric Administration Regional Climate Centers, Northeast Regional Climate Center at Cornell University for kindly sharing the meteorological data. We are thankful to the Environmental Protection Agency and the Department of Health, New York State for publishing the air quality data and daily infected records online.

\bibliography{references}
\bibliographystyle{apalike2}

\end{document}